\documentclass[12pt]{article}           
\usepackage{amsmath}
\usepackage{amssymb}
\usepackage[mathscr]{eucal}
\usepackage{epsfig}
\usepackage{eufrak}
\def\VR#1#2{\vrule height #1mm depth #2mm width 0pt}
\def\TVR#1#2{@{~~\VR{#1}{#2}}}
\newcommand{\por}[1]{\mbox{\boldmath${#1}$}}

\def\plb#1 #2 {Phys. Lett. {\bf #1B} #2 }
\def\phr#1 #2 {Phys. Rep. {\bf  #1} #2 }        
\def\npb#1 #2 {Nucl. Phys. {\bf B#1} #2 }
\def\aph#1 #2 {Ann. Phys. {\bf #1} #2 }         
\def\jmp#1 #2 {J. Math. Phys. {\bf #1} #2 }
\def\jgp#1 #2 {J. Geom. Phys. {\bf #1} #2 }
\def\prd#1 #2 {Phys. Rev. {\bf D#1} #2 }
\def\prl#1 #2 {Phys. Rev. Lett. {\bf #1} #2 }
\def\rmp#1 #2 {Rev. Mod. Phys.  {\bf #1} #2 }
\def\zpc#1 {Z. Phys. {\bf #1C} }
\def\cmp#1 #2 {Commun. Math. Phys. {\bf #1} #2 }
\def\cqg#1 #2 {Class.Quant.Grav. {\bf #1} #2 }
\def\mpl#1 {Mod. Phys. Lett. {\bf A#1} }
\def\cpc#1 {Computer Phys. Commun. {\bf #1} }   
\def\ijmp#1 {Int. J. Mod. Phys. {\bf A#1} }
\def\ijmpC#1 {Int. J. Mod. Phys. {\bf C#1} }

\begin{document}

\begin{flushright}
hep-th/9802170
\end{flushright}

\vspace{-6mm}

\hfill
TUW--98--05

\vspace{15mm}

\begin{center}
{\huge\bf A Note on (0,2) Models on \\

\vspace{3mm}

Calabi-Yau Complete Intersections} \\

\end{center}

\vspace{12mm}
\begin{center} 
\renewcommand{\thefootnote}{\fnsymbol{footnote}}  
{\large Mahmoud Nikbakht--Tehrani
\footnote[2]{e-mail: \texttt{nikbakht@tph16.tuwien.ac.at}}}

\vskip 5mm
       Institut f\"{u}r Theoretische Physik, Technische Universit\"{a}t Wien\\
       Wiedner Hauptstra\ss{}e 8--10, A-1040 Wien, AUSTRIA
\end{center}

\vspace{12mm}

\begin{center}
\begin{minipage}{15cm}
\begin{center}
{\bf Abstract}
\end{center}
In the class of $\,(0,2)\,$ heterotic compactifications 
which has been constructed in the framework of gauged 
linear sigma models the Calabi-Yau varieties $\,X\,$ are
realized as complete intersections of hypersurfaces in toric
varieties $\,{\Bbb P}_{\scriptscriptstyle\Sigma}\,$ and the 
corresponding gauge bundles (or more generally gauge sheaves) 
$\,\por{\mathscr E}\,$ are defined by some short exact sequences.
We show that there is yet another degree of freedom in resolving 
singularities in such models which is related to the possible choices
of nef partitions of the anticanonical divisors in Gorenstein 
Fano toric varieties ${\Bbb P}_{\scriptscriptstyle\Sigma}$.  
\end{minipage}
\end{center}

\vspace{15mm}

February  1998
\thispagestyle{empty}


\newpage


\section{Introduction}
\hrule
\vspace{4mm}

The gauged linear sigma model (GLSM) approach to the 
construction of $\,(0,2)\,$ heterotic string vacua \cite{wi93} 
has enabled a rather explicit study of the moduli space of 
this class of string vacua. This is  mainly due to the connection 
of this approach with the more accessible methods of toric 
geometry \cite{asp94}. 
Apart from the fact that the study of $\,(0,2)\,$ string 
vacua is per se of great interest, it is also desirable in 
other respects such as the duality between heterotic 
strings and F-theory compactifications \cite{fri97,ber97}. 
Recent years have witnessed some exciting progress in our 
understanding of the structure of the space of string vacua 
in which  such dualities play an important role. 

As a naive phase analysis of $\,(0,2)\,$ linear sigma 
models shows, in the geometrical phase the resulting Calabi-Yau
varieties are generally singular, signaling that this naive phase
picture is not complete. Therefore, we are led to resolve these 
singularities. The issue of resolving singularities in $\,(0,2)\,$ 
models has some peculiarities which are not shared by their 
better understood $\,(2,2)\,$ relatives. A physically sensible 
procedure for resolving singularities in $\,(0,2)\,$ models 
has been proposed in \cite{di96}. A more sophesticted analysis 
of some issues arising in this connection can be found in the 
recent works \cite{ber97,di97}.

Our purpose in this letter is to show that for $\,(0,2)\,$ models
whose target spaces are complete intersection Calabi-Yaus
in a toric variety $\,{\Bbb P}_{\scriptscriptstyle\Sigma}\,$ 
the choice of a nef partition represents yet another degree of
freedom which should also be taken into consideration.

The organization of this work is as follows. In the next section
we quickly review those aspects of GLSMs that we will need later. 
Section \ref{toric} provides the necessary mathematical background
from toric geometry. In section \ref{example} we bring an example
which explicitely demonstrates the role of nef partitions in the 
desingularization process. We conclude with some comments about
open problems.


\section{The gauged linear sigma models}
\hrule
\vspace{4mm}

In this section we briefly explain the basic ideas behind the 
GLSM approach without going into details. 
We will do this using a typical situation which we are interested 
in (cf. \cite{wi93,di96} for more details). 

The starting point is a $(0,2)$ supersymmetric $\,U(1)\,$ gauge 
theory that represents a nonconformal member of the universality 
class of a $(0,2)$ superconformal field theory. The action $\,S\,$ 
which describes  such a model is $\,S=S_{\mbox{\tiny gauge}}+
S_{\mbox{\tiny matter}}+S_{\scriptscriptstyle\mathscr W}+
S_{\scriptscriptstyle D,\theta}\,,$ where $\,S_{\mbox{\tiny gauge}}
\,$, $\,S_{\mbox{\tiny matter}}\,$ are the kinetic terms of the gauge
and matter fields, respectively, $\,S_{\scriptscriptstyle\mathscr W}\,$ 
is the superpotential and $\,S_{\scriptscriptstyle D,\theta}\,$ 
is the Fayet-Iliopoulos $D$-term and the $\,\theta$-angle term.
Let $\,P\,,\, \Phi_{1},\ldots, \Phi_{6}\,$ be chiral scalar superfields 
with $\,U(1)\,$ charges $\,-m, w_{1},\ldots, w_{6}\,$, and let 
$\,\Gamma^1$, $\Gamma^2, \Lambda^{1},\ldots, \Lambda^{5}\,$ be 
chiral spinor superfields with $\,U(1)\,$ charges $\,-d_1, 
-d_2, q_1,\ldots,q_5\,$ such that $\,m=\sum\nolimits_a q_a\,$
and $\,\sum\nolimits_j d_j=\sum\nolimits_i w_i\,$. 
The superpotential is given by 
\begin{eqnarray}
   S_{\scriptscriptstyle\mathscr W}=\int d^{2}z\, d\theta^{+}\,
   \left(\Gamma^j W_j(\Phi_{i})+P\Lambda^{a}F_{a}(\Phi_{i})\right)
   + h.c.\,,\label{super}
\end{eqnarray}
where $\,W_j\,$ and $\,F_a\,$ are homogeneous polynomials in 
$\,\Phi_{i}\,$ of degrees $d_j$ and $m-q_{a}$, respectively.
It is assumed that the $\,W_j\,$'s are transversal polynomials 
and that the $\,F_a\,$'s do not vanish simultaneously on 
$\,W_1(\phi_i)=W_2(\phi_i)=0\,$.
Integrating out the auxiliary $\,D\,$ field in the gauge multiplet and the 
auxiliary fields in the chiral spinor superfields, we get the scalar
potential
\begin{eqnarray*}
      U=\sum\nolimits_j|W_j(\phi_{i})|^{2}+|p|^{2}\;
      \sum\nolimits_{a}|F_{a}(\phi_{i})|^{2}+
      \frac{e^{2}}{2}\left(\sum\nolimits_{i}w_{i}|\phi_{i}|^{2}-m|p|^{2}
      -r\right)^{2}\,,
\end{eqnarray*}
where the parameter $\,r\,$ is the coefficient in the   
Fayet-Iliopoulos $D$-term and $\,\phi_{i}\,,\, p\,$ denote the 
lowest terms of the superfields $\,\Phi_{i}\,$ and $\,P\,$,
respectively. Now varying the parameter $\,r\,$ this model 
exhibits different `phases'. By minimizing the scalar potential 
$\,U\,$ for large positive values of  $\,r\,$ we obtain 
\begin{eqnarray*}
  \sum\nolimits_{i}w_{i}|\phi_{i}|^{2}=r\,,\;\;\; 
  W_1(\phi_{i})=W_2(\phi_{i})=0\,,\;\;\; p=0\; .
\end{eqnarray*}
Taking the quotient by the action of the $\,U(1)\,$ gauge group  
these equations describe a complete intersection Calabi-Yau 
variety $\,X\,$ as the zero locus of the homogeneous polynomials
$\,W_j(\phi_{i})\,$ in the weighted projective space 
$\,{\Bbb P}(w_{1},\ldots ,w_{6})\,$ with the K\"ahler class 
proportional to $\,r\,$. The right-moving fermions $\,\psi_{i}\,$ 
(the superpartners of $\phi_{i}$) couple to the tangent bundle of 
$\,X\,$. The left-moving fermions $\,\lambda^{a}\,$ 
(the lowest components of the spinor superfields $\,\Lambda^{a}\,$) 
couple to the vector bundle $\,V\,$ defined by the following exact
sequence 
\begin{eqnarray} 
   0\to V \to \bigoplus\nolimits_{a=1}^{5}{\cal O}(q_{a})
   \stackrel{F}{\longrightarrow}{\cal O}(m)\to 0\,.\label{bundle} 
\end{eqnarray}
So we find that our gauged linear sigma model for large positive 
values of $\,r\,$ reduces in the infrared limit to a $(0,2)$ 
Calabi-Yau $\sigma$ model with the target space $\,X\,$ being a complete
intersection of hypersurfaces in the weighted projective space 
$\,{\Bbb P}(w_{1},\ldots ,w_{6})\,$, and with a rank $4$ gauge bundle 
$\,V\,$ defined by (\ref{bundle}).
These geometric data still have to satisfy the following 
condition that comes from cancellation of 
the $\,U(1)\,$ gauge anomaly. 
This leads to the quadratic Diophantine equation:
\begin{equation*}
      m^{2}-\sum\nolimits_a q_{a}^{2}=\sum\nolimits_jd_j^{2}
      -\sum\nolimits_iw_{i}^{2}\;\;\label{anomaly}.
\end{equation*} 
It should be noted that the complete intersection Calabi-Yau 
varieties in weighted projective spaces are in general singular
\footnote{We assume that the Calabi-Yau varieties that arise
in this way have at worst canonical singularities. We recall 
that a (normal) variety $\,X\,$ is said to have canonical 
singularities if $\,mK_{X}\,$ is a Cartier divisor for some integer
$\,m\geq 1\,$ and if  $\,f:\tilde{X}\to X\,$ is a local resolution of
singularities then $\, mK_{\tilde{X}}=f^{*}(mK_{X})\,+\,\sum_{j}a_{j}
\,E_{j}\,$, where $\,K_{X}\,$ and $\,K_{\tilde{X}}\,$ are the 
canonical divisors of $\,X\,$ and $\,\tilde{X}\,$, respectively, 
$\,E_{j}\,$ are the exceptional prime divisors of $\,f\,$ and 
$\,a_{j}\,$ are nonnegative integers.}. 

In the process of desingularization we have to handle
two sets of data. At first we have to resolve the 
singularities of the base variety $\,X\,$. In the toric 
geometrical setting the starting point is a reflexive 
polytope $\,\Delta\,$ in a rank five lattice $\,{\bf N}\,$. 
This polytope defines a toric variety 
$\,{\Bbb P}_{\scriptscriptstyle\Sigma}\,$
which is generally a blowup of the weighted projective space 
$\,{\Bbb P}(w_{1},\ldots,w_{6})\,$. 
By taking a maximal triangulation of $\,\Delta\,$  we arrive
at a Calabi-Yau phase of the underlying model.
Note that a  maximal triangulation of $\,\Delta\,$ amounts 
above all to adding new 
one-dimensional cones to $\,\Sigma^{(1)}\,$ (= the set of one-dimensional
cones in $\,\Sigma\,$) which are associated with the 
points on the faces of $\,\Delta\,$\footnote{This is a consequence 
of our assumption on the type of singularities of $\,X\,$.}. 
In the context of gauged linear sigma
models these correspond to additional chiral scalar superfields
and additional $\,U(1)\,$ factors in the gauge group.
We also need to determine the charges of the fields with respect to 
the full gauge group. Translated into the
geometric language this means that we have to determine the degrees of
the variables in the homogeneous coordinate ring $\,S\,$. 
Let $\,x_{i}\,$ and $\,D_{e_i}\,$ denote the variables in the homogeneous 
coordinate ring $\,S\,$ \cite{cox92} and the divisors
associated to $\,e_{i}\,$ (= the primitive lattice vector on 
$\,\rho_i\in \Sigma^{(1)}\,$), respectively. The calculation of the 
cokernel of the map 
\begin{eqnarray*}
{\por\alpha}: {\bf M}\longrightarrow 
\bigoplus\nolimits_i{\Bbb Z}\cdot D_{e_i}\; ,\; m\mapsto 
\sum\nolimits_i\langle m,e_{i}\rangle \; D_{e_i}\,,
\end{eqnarray*}
where $\,{\bf M}= \mbox{Hom}({\bf N},{\Bbb Z})\,$ is  the dual
lattice, yields the desired quantities.
Note that the desingularization of the base variety simultaneously 
resolves the tangent sheaf to which the right-handed fermions couple. 
Therefore, these fermions have the same charges as their superpartners.
We still have to deal with the gauge bundle. 
Following \cite{di96} we take the solutions of the following
system of Diophantine equations 
\begin{eqnarray}
  m^{(k)}=\sum\nolimits_a q^{(k)}_{a}\hspace{4mm},\hspace{4mm}
  m^{(k)}m^{(\ell)}-\sum\nolimits_a q_{a}^{(k)}q_{a}^{(\ell)}=
  \sum\nolimits_jd_j^{(k)}d_j^{(\ell)}-\sum\nolimits_i
  w_{i}^{(k)}w_{i}^{(\ell)}\label{anomaly2}
\end{eqnarray}
with $\,\sum\nolimits_j d^{(k)}_j=\sum\nolimits_i w^{(k)}_{i}\,$ 
as possible gauge bundle data for the desingularized 
theory\footnote{$\,k\,,\,\ell = 1,\dots,\# U(1)\,$factors.}. As
we will show in section \ref{example}, the system (\ref{anomaly2}) 
depends in the case of complete intersection Calabi-Yau $\,X\,$ on 
the choice of a nef partition of $\,\Delta\,$.


\section{Calabi--Yau complete intersections and nef partitions}\label{toric}
\hrule
\vspace{4mm}

Let $\,{\bf N}\,$ and $\,{\bf M}=\mbox{Hom}({\bf N},{\Bbb Z})\,$
denote a dual pair of lattices of rank $\,d\,$ and $\,\Delta\,$ be a 
reflexive polytope in $\,{\bf N}_{\Bbb R}={\bf N}\otimes_{\Bbb Z}
{\Bbb R}\,$. Let $\,E=\{e_1,\ldots,e_n\}\,$ be the set of vertices 
of $\,\Delta\,$. Note that the $\,e_i\,$'s are primitive lattice 
vectors. Further, let $\,E=E_1\cup \ldots \cup E_r\,$ be a partition 
of $\,E\,$ and $\,D_i=\sum_{e_k\in E_i}\,D_{e_k}\,$, where 
$\,D_{e_k}\,$ denotes the $T$-invariant Weil divisor 
associated to the one-dimensional cone\footnote{It is the closed 
subvariety $\,X_{\mbox{\scriptsize cospan}\,\rho_k^\vee}\,$ in
$\,X_{\rho_k^\vee}\,$\cite{DA78}.} $\,\rho_k=
\langle\,e_k\,\rangle\,$. Recall that a divisor $\,D=\sum_j\,a_j
D_{e_j}\,$ is a $T$-invariant Cartier divisor if and only if there 
exists a continuous real function $\,\psi_D\,$ on
$\,|\Sigma |\,$, the support of the fan $\,\Sigma\,$, with 
$\,\psi_D(e_j)=a_j\,$ such that\vspace{-3mm} 
\begin{itemize}
\item $\,\psi_D\,$ is integral, i.e. $\,\psi_D(\,|\Sigma|\cap {\bf
      N}\,)\subset {\Bbb Z}\,$,\\ \vspace{-8mm}
\item $\,\psi_D\,$ is $\,\Sigma$-piecewise linear, i.e. the restriction of \\
      $\,\psi_D\,$ to each cone in $\,\Sigma\,$ is an $\,\Bbb
      R$-linear function. 
\end{itemize}\vspace{-3mm}
If $\,\psi_D\,$ is convex,
i.e. $\,\psi_D(t{\por v}+(1-t){\por w})\leq t\psi_D({\por v})+(1-t)
\psi_D({\por w})\,$ for all $\,{\por v},{\por w}\in |\Sigma |\,$
and $\,t\in [0,1]\,$, then $\,D\,$ is semi-ample which means that
$\,{\cal O}_{{\Bbb P}_{\scriptscriptstyle\Sigma}}(D)\,$ is generated 
by its global sections.

A nef partition of $\,E\,$ is a partition $\,E=E_1\cup \ldots \cup
E_r\,$ such that its corresponding divisors $\,D_i\,$
($\,=\sum_{e_k\in E_i}\,D_{e_k}\,$) are semi-ample $T$-invariant 
Cartier divisors. This can be  equivalently formulated as follows.
Note that the anticanonical divisor $\,-K=\sum_{j=1}^{n}\,D_{e_j}\,$ of
the toric variety $\,{\Bbb P}_{\scriptscriptstyle\Sigma}\,$ 
constructed from $\,\Delta\,$ is an ample Cartier divisor. 
Therefore, its corresponding function $\,\por\psi\,$ is 
(strictly) convex. $\,E=E_1\cup \ldots \cup E_r\,$ is a nef
partition of $\,E\,$ if there exists $\,\Sigma$-piecewise linear 
integral convex functions $\,\psi_i\,$, $i=1,\ldots,r\,$, with
$\psi_i(e_k)=1\,$ for $\,e_k\in E_i\,$ and $\psi_i(e_k)=0\,$ for 
$\,e_k\not\in E_i\,$ such that $\,{\por\psi}=\psi_1+\ldots+\psi_r\,$. 
 
A complete intersection Calabi-Yau variety 
with only canonical singularities can be realized in a Gorenstein 
Fano toric variety in the following way. 
Using the correspondence of reflexive polytopes 
and Gorenstein  Fano toric varieties we begin first with a reflexive 
polytope $\,\Delta\,$ in $\,{\bf N}\,$ and construct its corresponding
toric variety $\,{\Bbb P}_{\scriptscriptstyle\Sigma}\,$. Next we 
consider a nef partition of the anticanonical divisor 
$\,-K=\sum^{r}_{i=1}D_{i}$ of $\,{\Bbb P}_{\scriptscriptstyle\Sigma}\,$. 
Now let $Y_{i}$ be a generic section of 
${\cal O}_{{\Bbb P}_{\scriptscriptstyle\Sigma}}(D_{i})$. 
Then the complete intersection $\,\bigcap^{r}_{i=1}Y_{i}\,$ 
will be a canonical Calabi-Yau variety of codimension $\,r\,$ in 
$\,{\Bbb P}_{\scriptscriptstyle\Sigma}\,$ \cite{bor,bat}.


\section{An example in detail}\label{example}
\hrule
\vspace{4mm}

We begin with the superpotentail (\ref{super}) with
\begin{center}
\begin{tabular}{c||c|c|c|c|c|c|c||c|c|c|c|c|c|c\TVR50} 
\itshape{\bfseries field}  & $P$ & $\Phi_1$ & $\Phi_2$ &$\Phi_3$&
                             $\Phi_4$&$\Phi_5$&$\Phi_6$   
                            & $\Gamma^1$ & $\Gamma^2$ &$\Lambda^1$ & 
                             $\Lambda^2$ &$\Lambda^3$&$\Lambda^4$&
                             $\Lambda^5$ \\\hline   
\itshape{\bfseries charge} &$-6$&$1$&$1$&$1$&
                            $1$&$2$&$4$
                           &$-6$&$-4$&$1$&$1$&$1$&
                            $1$&$2$\\
\end{tabular}
\end{center}\vspace{-6mm} 
\begin{center}
\begin{tabular}{c||c|c|c|c|c|c|c\TVR50} 
\itshape{\bfseries poly. }  & $W_1$ & $W_2$ &$F_1$ & 
                             $F_2$ &$F_3$&$F_4$&
                             $F_5$   \\\hline
\itshape{\bfseries deg.} &$6$&$4$&$5$&$5$&$5$&
                            $5$&$4$\\
\end{tabular}
\end{center} 
which reduces in its `Calabi-Yau phase' to a $(0,2)$ sigma model
with target space $\,X\,$ a codimension $2$ complete intersection 
Calabi-Yau variety in $\,{\Bbb P}(1,1,1,1,2,4)\,$. The gauge bundle
$\,V\,$ is defined by the following short exact sequence 
\begin{eqnarray} 
   0\to V \to {\cal O}(1)^{\oplus 4}\oplus {\cal O}(2)
   \stackrel{F}{\longrightarrow}{\cal O}(6)\to 0\,.\label{bundle2} 
\end{eqnarray}
The reflexive polytope $\Delta$ corresponding to 
${\Bbb P}(1,1,1,1,2,4)$ is given by
\begin{eqnarray*}
\begin{array}{llll}
&e_{1}=(1,0,0,0,0)&e_{2}=(0,1,0,0,0)&e_{3}=(0,0,1,0,0)\hspace{3mm}
                                     e_{4}=(0,0,0,1,0)\\
&e_{5}=(0,0,0,0,1)&e_{6}=(0,0,0,0,-1)&e_{7}=(-1,-1,-1,-2,-4)\,.
\end{array}
\end{eqnarray*}
with respect to the canonical basis in the rank five lattice 
$\,{\bf N}\,$. It can be easily checked that there are no other 
points on the faces of $\,\Delta\,$. Taking a maximal triangulation 
of the reflexive polytope $\,\Delta\,$ leads to a simplicial fan 
$\,\Sigma\,$ whose big cones are defined by
\begin{eqnarray*}
\begin{array}{llll}
   \sigma_{1}=\langle e_{1}e_{2}e_{3}e_{4}e_{5}\rangle &
   \sigma_{2}=\langle e_{1}e_{2}e_{3}e_{4}e_{6}\rangle &
   \sigma_{3}=\langle e_{1}e_{2}e_{3}e_{5}e_{7}\rangle &
   \sigma_{4}=\langle e_{1}e_{2}e_{3}e_{6}e_{7}\rangle \\
   \sigma_{5}=\langle e_{1}e_{2}e_{4}e_{5}e_{7}\rangle &
   \sigma_{6}=\langle e_{1}e_{2}e_{4}e_{6}e_{7}\rangle &
   \sigma_{7}=\langle e_{1}e_{3}e_{4}e_{5}e_{7}\rangle &
   \sigma_{8}=\langle e_{1}e_{3}e_{4}e_{6}e_{7}\rangle \\
   \sigma_{9}=\langle e_{2}e_{3}e_{4}e_{5}e_{7}\rangle &
   \sigma_{10}=\langle e_{2}e_{3}e_{4}e_{6}e_{7}\rangle &&\\
\end{array}
\end{eqnarray*}
The Fano toric variety $\,{\Bbb P}_{\scriptscriptstyle\Sigma}\,$ 
constructed from the simplicial fan $\,\Sigma\,$ is a blowup 
of the weighted projective space $\,{\Bbb P}(1,1,1,1,2,4)\,$. 
The one-dimensional cone $\langle e_{6}\rangle$ corresponds to 
the resulting exceptional divisor. The variety 
${\Bbb P}_{\scriptscriptstyle\Sigma}\,$ still has 
a curve of $\,{\Bbb Z}_2$--cyclic quotient singularities. 
Let $\,x_{i}\,$ 
denote the variables in the homogeneous coordinate ring $\,S\,$
which correspond to the bosonic fields of the model 
under consideration. 
The calculation of the 
cokernel of the map $\,{\por\alpha}\,$
in our case yields
\begin{center}
\begin{tabular}{c||c|c|c|c|c|c|c\TVR50} 
\itshape{\bfseries field}  & $x_1$ & $x_2$ &$x_3$& $x_4$&$x_5$&
                             $x_6$ & $x_7$  \\\hline
\itshape{\bfseries charge} &$(1,0)$&$(1,0)$&$(1,0)$&$(2,0)$&
                            $(4,1)$&$(0,1)$&$(1,0)$\\
\end{tabular}
\end{center} 

\subsection*{The nef partitions}

Let $\,X_1,\ldots,X_5\,$ denote the coordinate functions on
$\,{\bf N}_{\Bbb R}\,$ with respect to the canonical basis and
$\varphi(X_1,\ldots,X_5)=a_1X_1+\ldots +a_5X_5\,$ be a typical
linear form on it. Each integer solution of the following systems
of equations, in which $\,\varphi_i(e_j)\in {\Bbb Z}\,$, gives us a 
($T$-invariant) Cartier divisor: 
{\small
\begin{center}
\begin{tabular}{c|c} 
$\begin{array}{c}
\begin{array}{ll}
\fbox{$\sigma_1$}\leftrightarrow\varphi_1
  &a_1=\varphi_1(e_1)\\
  &a_2=\varphi_1(e_2)\\
  &a_3=\varphi_1(e_3)\\
  &a_4=\varphi_1(e_4)\\
  &a_5=\varphi_1(e_5)
\end{array}
\\
\end{array}$&
$\begin{array}{c}
\begin{array}{ll}
\fbox{$\sigma_2$}\leftrightarrow\varphi_2
  &a_1=\varphi_2(e_1)\\
  &a_2=\varphi_2(e_2)\\
  &a_3=\varphi_2(e_3)\\
  &a_4=\varphi_2(e_4)\\
  &-a_5=\varphi_2(e_6)  
\end{array}
\\
\end{array}$\\\hline
$\begin{array}{c}
\begin{array}{ll}
\fbox{$\sigma_3$}\leftrightarrow\varphi_3
  &a_1=\varphi_3(e_1)\\
  &a_2=\varphi_3(e_2)\\
  &a_3=\varphi_3(e_3)\\
  &a_5=\varphi_3(e_5)\\
\end{array}\\
   -a_1-a_2-a_3-2a_4-4a_5=\varphi_3(e_7)\\ 
                          \\
\end{array}$&
$\begin{array}{c}
\\
\begin{array}{ll}
\fbox{$\sigma_4$}\leftrightarrow\varphi_4
  &a_1=\varphi_4(e_1)\\
  &a_2=\varphi_4(e_2)\\
  &a_3=\varphi_4(e_3)\\
  &-a_5=\varphi_4(e_6)\\
\end{array}\\
  -a_1-a_2-a_3-2a_4-4a_5=\varphi_4(e_7) \\
                          \\
\end{array}$\\\hline
$\begin{array}{c}
\\
\begin{array}{ll}
\fbox{$\sigma_5$}\leftrightarrow\varphi_5
  &a_1=\varphi_5(e_1)\\
  &a_2=\varphi_5(e_2)\\
  &a_4=\varphi_5(e_4)\\
  &a_5=\varphi_5(e_5)\\
\end{array}\\
  -a_1-a_2-a_3-2a_4-4a_5=\varphi_5(e_7) \\
                          \\  
\end{array}$&
$\begin{array}{c}
\\
\begin{array}{ll}
\fbox{$\sigma_6$}\leftrightarrow\varphi_6
  &a_1=\varphi_6(e_1)\\
  &a_2=\varphi_6(e_2)\\
  &a_4=\varphi_6(e_4)\\
  &-a_5=\varphi_6(e_6)\\
\end{array}\\
  -a_1-a_2-a_3-2a_4-4a_5=\varphi_6(e_7) \\
                          \\
\end{array}$\\\hline
$\begin{array}{c}
\\
\begin{array}{ll}
\fbox{$\sigma_7$}\leftrightarrow\varphi_7
  &a_1=\varphi_7(e_1)\\
  &a_3=\varphi_7(e_3)\\
  &a_4=\varphi_7(e_4)\\
  &a_5=\varphi_7(e_5)\\
\end{array}\\
  -a_1-a_2-a_3-2a_4-4a_5=\varphi_7(e_7)  \\
                                         \\
\end{array}$&
$\begin{array}{c}
\\
\begin{array}{ll}
\fbox{$\sigma_8$}\leftrightarrow\varphi_8
  &a_1=\varphi_8(e_1)\\
  &a_3=\varphi_8(e_3)\\
  &a_4=\varphi_8(e_4)\\
  &-a_5=\varphi_8(e_6)\\
\end{array}\\
  -a_1-a_2-a_3-2a_4-4a_5=\varphi_8(e_7) \\
                                        \\
\end{array}$\\\hline
$\begin{array}{c}
\\
\begin{array}{ll}
\fbox{$\sigma_9$}\leftrightarrow\varphi_9
  &a_2=\varphi_9(e_2)\\
  &a_3=\varphi_9(e_3)\\
  &a_4=\varphi_9(e_4)\\
  &a_5=\varphi_9(e_5)\\
\end{array}\\
  -a_1-a_2-a_3-2a_4-4a_5=\varphi_9(e_7) \\
                                        \\
\end{array}$&
$\begin{array}{c}
\begin{array}{ll}
\fbox{$\sigma_{10}$}\leftrightarrow\varphi_{10}
  &a_2=\varphi_{10}(e_2)\\
  &a_3=\varphi_{10}(e_3)\\
  &a_4=\varphi_{10}(e_4)\\
  &-a_5=\varphi_{10}(e_6)\\
\end{array}\\
  -a_1-a_2-a_3-2a_4-4a_5=\varphi_{10}(e_7) 
\end{array}$ 
\end{tabular}
\end{center} 
}
$\fbox{\bf 1}\,$ For the partition $\,\{e_1,e_2,e_3,e_4,e_6,e_7\}
\cup\{e_5\}\,$ we see that the corresponding divisors $\,D_1\,$ 
and $\,D_2\,$ are indeed Cartier. 
Therefore, $\,D_1=\sum_{j}\,{\por\varphi}_1(e_j)D_{e_j}\,$ with 
$\,{\por\varphi}_1(e_j)=1\,$ for $\,e_j\in E_1\,$ and 
$\,{\por\varphi}_1(e_j)=0\,$
otherwise, and $\,D_2=\sum_{j}\,{\por\varphi}_2(e_j)D_{e_j}\,$ with
$\,{\por\varphi}_2(e_j)=0\,$ for $\,e_j\in E_1\,$ and 
$\,{\por\varphi}_2(e_j)=1\,$
otherwise. The functions $\,{\por\varphi}_1\,$ and 
$\,{\por\varphi}_2\,$ are 
given by\footnote{$\varphi_i$ denotes the restriction of both
$\,{\por\varphi}_1\,$ and $\,{\por\varphi}_2\,$ to the $i$-the big cone of 
$\,\Sigma\,$.} 
\begin{center}
\begin{tabular}{ll}
$\varphi_1(X_1,\ldots,X_5)=X_1+X_2+X_3+X_4 $    &\hspace{4mm}
$\varphi_1(X_1,\ldots,X_5)=X_5 $                \\
$\varphi_2(X_1,\ldots,X_5)=X_1+X_2+X_3+X_4-X_5$ &\hspace{4mm}
$\varphi_2(X_1,\ldots,X_5)=0$                   \\
$\varphi_3(X_1,\ldots,X_5)=X_1+X_2+X_3-2X_4$    &\hspace{4mm}
$\varphi_3(X_1,\ldots,X_5)=-2X_4+X_5 $          \\
$\varphi_4(X_1,\ldots,X_5)= X_1+X_2+X_3-X_5$    &\hspace{4mm}
$\varphi_4(X_1,\ldots,X_5)=0 $                  \\
$\varphi_5(X_1,\ldots,X_5)=X_1+X_2-5X_3+X_4 $   &\hspace{4mm}
$\varphi_5(X_1,\ldots,X_5)=-4X_3+X_5 $          \\
$\varphi_6(X_1,\ldots,X_5)=X_1+X_2+X_4-X_5 $    &\hspace{4mm}
$\varphi_6(X_1,\ldots,X_5)=0 $                  \\
$\varphi_7(X_1,\ldots,X_5)=X_1-5X_2+X_3+X_4 $   &\hspace{4mm}
$\varphi_7(X_1,\ldots,X_5)=-4X_2+X_5 $          \\
$\varphi_8(X_1,\ldots,X_5)=X_1+X_3+X_4-X_5 $    &\hspace{4mm}
$\varphi_8(X_1,\ldots,X_5)=0 $                  \\
$\varphi_9(X_1,\ldots,X_5)=-5X_1+X_2+X_3+X_4 $  &\hspace{4mm}
$\varphi_9(X_1,\ldots,X_5)=-4X_1+X_5 $          \\
$\varphi_{10}(X_1,\ldots,X_5)=X_2+X_3+X_4-X_5 $ &\hspace{4mm}
$\varphi_{10}(X_1,\ldots,X_5)=0 $
\end{tabular}
\end{center} 
After some elementary but tedious calculations we find out that
$\,{\por\varphi}_1\,$ and $\,{\por\varphi}_2\,$ are convex, which 
means that the above partition is nef \footnote{This nef partition 
corresponds to a Calabi-Yau complete intersection of hypersurfaces 
of degrees $\,{\por d}_1=(6,0)\,$ and $\,{\por d}_2=(4,1)\,$, 
respectively.}.

\noindent
$\fbox{\bf 2}\,$ For the partition $\,\{e_1,e_2,e_3,e_6,e_7\}\cup
\{e_4,e_5\}\,$ we find that the corresponding divisors 
$\,D_1\,$ and $\,D_2\,$ are also Cartier. Therefore, 
$\,D_1=\sum_{j}\,{\por\varphi}_1(e_j)D_{e_j}\,$ with 
$\,{\por\varphi}_1(e_j)=1\,$ for $\,e_j\in E_1\,$ and 
$\,{\por\varphi}_1(e_j)=0\,$
otherwise, and $\,D_2=\sum_{j}\,{\por\varphi}_2(e_j)D_{e_j}\,$ with
$\,{\por\varphi}_2(e_j)=0\,$ for $\,e_j\in E_1\,$ and 
$\,{\por\varphi}_2(e_j)=1\,$
otherwise. The functions $\,{\por\varphi}_1\,$ and 
$\,{\por\varphi}_2\,$ are 
given by
\begin{center}
\begin{tabular}{ll}
$\varphi_1(X_1,\ldots,X_5)=X_1+X_2+X_3 $     &\hspace{8mm}
$\varphi_1(X_1,\ldots,X_5)=X_4+X_5 $         \\
$\varphi_2(X_1,\ldots,X_5)=X_1+X_2+X_3-X_5$  &\hspace{8mm}
$\varphi_2(X_1,\ldots,X_5)=X_4$              \\
$\varphi_3(X_1,\ldots,X_5)=X_1+X_2+X_3-2X_4$ &\hspace{8mm}
$\varphi_3(X_1,\ldots,X_5)=-2X_4+X_5 $       \\
$\varphi_4(X_1,\ldots,X_5)= X_1+X_2+X_3-X_5$ &\hspace{8mm}
$\varphi_4(X_1,\ldots,X_5)=0 $               \\
$\varphi_5(X_1,\ldots,X_5)=X_1+X_2-3X_3$     &\hspace{8mm}
$\varphi_5(X_1,\ldots,X_5)=-6X_3+X_4+X_5 $   \\
$\varphi_6(X_1,\ldots,X_5)=X_1+X_2+X_3-X_5 $ &\hspace{8mm}
$\varphi_6(X_1,\ldots,X_5)=-2X_3+X_4$        \\
$\varphi_7(X_1,\ldots,X_5)=X_1-3X_2+X_3 $    &\hspace{8mm}
$\varphi_7(X_1,\ldots,X_5)=-6X_2+X_4+X_5 $   \\
$\varphi_8(X_1,\ldots,X_5)=X_1+X_2+X_3-X_5 $ &\hspace{8mm}
$\varphi_8(X_1,\ldots,X_5)=-2X_2+X_4 $       \\
$\varphi_9(X_1,\ldots,X_5)=-3X_1+X_2+X_3 $   &\hspace{8mm}
$\varphi_9(X_1,\ldots,X_5)=-6X_1+X_4+X_5 $   \\
$\varphi_{10}(X_1,\ldots,X_5)=X_1+X_2+X_3-X_5$ &\hspace{8mm}
$\varphi_{10}(X_1,\ldots,X_5)=-2X_1+X_4 $
\end{tabular}
\end{center} 
$\,{\por\varphi}_1\,$ and $\,{\por\varphi}_2\,$ are also convex.
Therefore, the above partition is also nef \footnote{This nef 
partition corresponds to a Calabi-Yau complete intersection of 
hypersurfaces of degrees $\,{\por d}_1=(6,1)\,$ and 
$\,{\por d}_2=(4,0)\,$, respectively.}.

The data of gauge bundle in the desingularized theory can be determined  
for the first nef partition from the following system of Diophantine 
equations (cf. (\ref{anomaly2}))
\begin{eqnarray}
&&q_1+q_2+q_3+q_4+q_5-p=0\nonumber\\
&&q_1+q_2+q_3+q_4+2q_5-6p=0\label{par1}\\
&&q^2_1+q^2_2+q^2_3+q^2_4+q^2_5-p^2=0\nonumber\,.
\end{eqnarray}
For the second partition the same data can be determined from
the system 
\begin{eqnarray}
&&q_1+q_2+q_3+q_4+q_5-p=0\nonumber\\
&&q_1+q_2+q_3+q_4+2q_5-6p=-2\label{par2}\\
&&q^2_1+q^2_2+q^2_3+q^2_4+q^2_5-p^2=0\nonumber\,.
\end{eqnarray}
It can be easily seen  that the trivial solution
$\,q_1=\ldots=q_5=p=0\,$ is the only solution of (\ref{par1})
while the system (\ref{par2}) doesn't have any solution which
means that the model whose target variety is defined by the second
nef partition does not admit a desingularization!

\subsection*{The Euler characteristic}

$\underline{\mbox{\bf The nonsingular model}\; \fbox{\bf 1}}$

The primitive collections of $\,\Sigma\,$ are
$\,\{e_5,e_6\},\{e_1,e_2,e_3,e_4,e_7\}\,$. 
Taking the quotient of intersection ring $\,A^{\bullet}({\Bbb P}
_{\scriptscriptstyle\Sigma})_{\Bbb Q}={\Bbb Q}[x_1,\ldots,x_8]/\,I+J\,$
of $\,{\Bbb P}_{\scriptscriptstyle\Sigma}\,$, where
\begin{eqnarray*}  
I=\langle\,x_1-x_7\,,\, x_2-x_7\,,\,x_3-x_7\,,\,
x_4-2x_7\,,\, x_5-x_6-4x_7\,\rangle\hspace{2mm}, \hspace{2mm}
J=\langle\,x_5x_6\,,\, x_1x_2x_3x_4x_7\,\rangle\,,
\end{eqnarray*}
by the annihilator of $\,x_5\cdot (x_1+x_2+x_3+x_4+x_6+x_7)\,$ we arrive
at $\,A^{\bullet}(\tilde{X})_{\Bbb Q}\,$. The calculation of
the third Chern class yields 
$\,c_3(\tilde{V})= -68\,x_7^3\,$ in $A^{\bullet}(\tilde{X})_{\Bbb Q}\,$. 
Since $\,\langle\,D_1D_2D_3D_5D_7\,\rangle=\langle\,D_1D_2D_3D_6D_7\,
\rangle=1/2\,$ and  $\,\langle\,D_iD_jD_kD_lD_m\,\rangle=1\,$ for all other
big cones we find by using the `algebraic moving lemma' 
that the normalization in $A^{\bullet}(\tilde{X})_{\Bbb Q}\,$ is
$\,\langle\,x_7^3  \,\rangle=5/2\,$. Therefore,
$\,\chi(\tilde{V})=-85\,$.

\noindent
$\underline{\mbox{\bf The singular model}\; \fbox{\bf 2}}$

Let $\,\frak Z\,$ and $\,\frak Y\,$ denote hypersurfaces 
of degrees $6$ and $4$, respectively, whose complete intersection
defines the Calabi-Yau variety $\,X\,$ in $\,{\Bbb P}$ (:=
${\Bbb P}(1,1,1,1,2,4)\,$)\footnote{$\,X\,$ is a well-formed complete
intersection 
which allows
us to use the adjunction formula \cite{flet}.}. Using 
\begin{eqnarray*}  
&&    0\to{\cal O}_{\Bbb P}(-6)\to{\cal O}_{\Bbb P}
       \to{\cal O}_{\frak Z}\to 0\,,\\ 
&&    0\to{\cal O}_{\frak Z}(-4)\to{\cal O}_{\frak Z}
       \to{\cal O}_{X}\to 0\,,
\end{eqnarray*}
and (\ref{bundle2}) together with  their corresponding long 
exact cohomology sequences and taking into account that \cite{cohom}
\begin{eqnarray*} 
   H^p({\Bbb P},{\cal O}_{\Bbb P}(q))=\left\{
\begin{array}{lll}
   \bigoplus\limits_{n_i\geq 0\atop {i=1,\ldots,6}}\!\!{\Bbb C}\cdot 
   x_1^{n_1}\!\ldots\; x_6^{n_6}\big|_{\mbox{\scriptsize\sl deg.}\;q}&
   \hspace{5mm}\mbox{for} & p=0\\\vspace{-3mm}
                   \\
   \hspace{5mm}0 & \hspace{5mm}\mbox{for} & 0<p<5\\
                         \\
   \bigoplus\limits_{n_i> 0\atop {i=1,\ldots,6}}\!\!\!{\Bbb C}\cdot 
   x_1^{-n_1}\!\!\ldots\; x_6^{-n_6}\big|_{\mbox{\scriptsize\sl deg.}\;q}&
   \hspace{5mm}\mbox{for} & p=5
\end{array}
\right.
\end{eqnarray*}
we obtain
\begin{center}
\begin{tabular}{l|l|l\TVR50} 
  $h^0(X,{\cal O}_{X}(1))=4$ &$h^0(X,{\cal O}_{X}(2))=11$ & 
  $h^0(X,{\cal O}_{X}(6))=129$\\\hline
  $h^3(X,{\cal O}_{X}(1))=0$ &$h^3(X,{\cal O}_{X}(2))=0$  &
  $h^3(X,{\cal O}_{X}(6))=0$\\
\end{tabular}
\end{center} 
which yields the result\footnote{Note that the stability
of gauge bundle also implies $\,H^0(X,V)=H^3(X,V)=0\,$!}
$\,h^1(X,V)=102\,$ and $\,h^i(X,V)=0\,$ for
$\,i\neq 1\,$. Therefore $\,\chi(V)=-102\,$.


\section{Conclusion}
\hrule
\vspace{4mm}

We have seen that if the target space of a $\,(0,2)\,$ model
is a complete intersection Calabi-Yau variety, then there appears
an additional degree of freedom in the desingularization process
which is related to the possible choices of nef partitions
of the defining reflexive polytope of the ambient toric variety.
Obviously, the only use of the anomaly cancellation conditions and
the above degree of freedom yields plenty of possible 
desingularized models in  most cases. A question which can 
be immediately posed is which of these possible desingularizations
are physically admissible vacua  and how are they related. 
It is expected that the stability of gauge bundles imposes severe
restrictions here, but not much about this is known.
There is a second point which is worthwhile a moment of 
reflection. We have seen that in some cases there is no
desingularization for a given $\,(0,2)\,$ model although 
the base Calabi-Yau variety is not minimal, a fact 
which contrasts the $\,(2,2)\,$ case, where the
existence of a minimal Calabi-Yau variety is guaranteed.
It would be interesting to know which physical consequences 
such an obstruction has.    

\vspace{6mm}

\noindent
{\bf Acknowledgement.}
I would like to thank M. Kreuzer for reading the manuscript 
and for useful comments. This work has been supported by the Austrian Research
Fund (FWF) under grant Nr. P10641-PHY and \"ONB under grant Nr. 6632.\\




\end{document}